\documentclass[twocolumn,secnumarabic,amssymb,nobibnotes,aps,prd,superscriptaddress]{revtex4-2}
\usepackage{booktabs}
\usepackage{float}
\usepackage{graphicx} 
\usepackage{subfigure}
\usepackage{multirow}
\usepackage{xspace}
\usepackage{fancyhdr}
\usepackage{amsmath,amssymb}
\usepackage{xcolor}
\usepackage{amsmath}
\usepackage{amssymb}
\usepackage{amstext}
\usepackage{amsopn}
\usepackage{amsfonts}
\usepackage{amsxtra}
\usepackage[english]{babel}
\usepackage{bm}
\usepackage{dcolumn}
\usepackage{color}
\usepackage{textcomp}
\usepackage{upgreek}
\usepackage{textgreek}
\usepackage{setspace}
\usepackage{booktabs}
\usepackage{makecell}
\usepackage{tabularx}
\usepackage{ragged2e}
\usepackage{array} 
\usepackage{tabularray}
\usepackage{threeparttable}
\usepackage{hyperref}
\hypersetup{colorlinks,breaklinks,
            urlcolor=[rgb]{0,0,0.64},
            linkcolor=[rgb]{0,0,0.64},
            citecolor=[rgb]{0,0,0.64}}
\usepackage{nccmath}
\usepackage{bm}
\usepackage{lineno}
\usepackage{siunitx}  

\begin{document}

\newcommand{\NJU}{National Laboratory of Solid State Microstructures and Department of Physics, Nanjing University, Nanjing 210093, China}
\newcommand{\IOP}{Institute of Physics, Chinese Academy of Sciences, Beijing 100190, China}
\newcommand{\CAS}{Beijing National Laboratory for Condensed Matter Physics, Institute of Physics, Chinese Academy of Sciences, Beijing 100190, China}
\newcommand{\JPSRC}{Jiangsu Physical Science Research Center, Nanjing University, Nanjing 210093, China}

\title{Finite-Momentum Antiferromagnetic Magnon Dynamics across the Morin Transition in Hematite}

\author{Jiaqi~Gong}
\thanks{These authors contributed equally to this work.}
\affiliation{\NJU}

\author{Yuanzhe~Tian}
\thanks{These authors contributed equally to this work.}
\affiliation{\NJU}

\author{Jun~Cui}
\thanks{These authors contributed equally to this work.}
\affiliation{\NJU}

\author{Meiye~Hou}
\affiliation{\NJU}

\author{Yuxuan~Mu}
\affiliation{\NJU}

\author{Di~Wang}
\affiliation{\NJU}
\affiliation{\JPSRC}

\author{Xiangang Wan}
\affiliation{\NJU}
\affiliation{\JPSRC}

\author{Di Wu}
\thanks{Correspondence to: \href{mailto:dwu@nju.edu.cn}{dwu@nju.edu.cn}}
\affiliation{\NJU}
\affiliation{\JPSRC}

\author{Qi Zhang}
\thanks{Correspondence to: \href{mailto:zhangqi@nju.edu.cn}{zhangqi@nju.edu.cn}}
\affiliation{\NJU}
\affiliation{\JPSRC}

\date{\today}

\begin{abstract}
Hematite ($\alpha$-Fe$_2$O$_3$) is a prototypical antiferromagnetic platform for high-speed spintronics and magnonics, yet its finite-momentum magnon dynamics and spin-reorientation mechanism remain incompletely understood. Here we use low-wavenumber magneto-Raman spectroscopy to resolve both $k=0$ and finite-$k$ sub-terahertz magnons in $\alpha$-Fe$_2$O$_3$ across the Morin transition. We find that the limiting group velocity $v_0$ remains nearly field independent, whereas the finite-$k$ group velocity $v_{\mathrm g}$ is strongly modified near the Morin and spin-flop transitions. By combining parallel- and transverse-field measurements with spin-wave modeling, we further extract the temperature-dependent Dzyaloshinskii--Moriya field $H_{\mathrm{D}}$ and uniaxial anisotropy fields $H_{\mathrm{K1}}$ and $H_{\mathrm{K2}}$. We find that $H_{\mathrm{D}}$ is weakly temperature dependent at approximately 2.0--2.3~T, while $H_{\mathrm{K1}}$ decreases rapidly and $H_{\mathrm{K2}}$ changes only slightly. This contrast drives the sign reversal of the effective anisotropy field and quantitatively accounts for the Morin transition. Our results establish low-wavenumber Raman spectroscopy as a quantitative probe of finite-momentum antiferromagnetic magnons and provide key material parameters for hematite-based magnonics.
\end{abstract}
\maketitle

\section{Introduction}   
Antiferromagnetic (AFM) magnons are attractive information carriers for spintronic technologies because they combine terahertz (THz) resonance frequencies, negligible stray fields, and fast dynamics with low dissipation~\cite{jungwirth2016antiferromagnetic,vaidya2020subterahertz,chen2024emerging,liu2025antiferromagnetic,fritjofson2025coherent,chumak2015magnon}. Among AFM materials, hematite ($\alpha$-Fe$_2$O$_3$) is a prototypical platform owing to its low damping~\cite{searle1968magnetic,lebrun2020long,boventer2023antiferromagnetic,lebrun2018tunable}, strong exchange interaction~\cite{hamdi2023spin,wang2023long,wang2024perspectives}, and Morin transition~\cite{ozdemir2008morin,jani2021reversible,kubaniova2019hematite}, which switches the system between easy-axis and easy-plane magnetic order. This spin-reorientation transition provides a unique setting for probing how magnetic anisotropy, the Dzyaloshinskii-Moriya (DM) interaction, and exchange jointly govern magnon dynamics across distinct magnetic phases~\cite{liu2025anisotropic,besser1967magnetocrystalline,morrish1994canted}.

Driven by the demand for next-generation spintronic devices, finite-$k$ magnons in hematite have recently attracted considerable attention. Propagating spin-wave spectroscopy~\cite{el2023antiferromagnetic,wang2023long}, Brillouin light scattering~\cite{hamdi2023spin}, pump-probe measurements~\cite{xu2026handed,li2026anisotropic,yang2024spin}, and electrical methods~\cite{lebrun2020long,han2020birefringence,xu2026optical} have revealed long-distance propagation, high group velocity, nonreciprocal transport, and tunability, highlighting hematite as a promising platform for ultrafast magnonics. However, a complete experimental picture of finite-$k$ sub-THz magnons across the Morin transition is still lacking. In particular, their field-dependent evolution under different magnetic-field orientations, as well as the temperature dependence of the limiting group velocity $v_0$ and the microscopic free-energy parameters that determine the spin-reorientation transition, remain insufficiently constrained.

Here, we use low-wavenumber Raman spectroscopy to directly resolve both $k=0$ and finite-$k$ magnons in $\alpha$-Fe$_2$O$_3$ across the Morin transition, and further identify the left- and right-handed finite-$k$ magnon modes in the low-temperature phase. By systematically mapping the magnon spectra under different temperatures and magnetic-field orientations, we determine the DM field $H_{\mathrm D}$ together with the first- and second-order uniaxial anisotropy fields $H_{\mathrm{K1}}$ and $H_{\mathrm{K2}}$. In particular, the joint analysis of parallel- and transverse-field spectra provides independent constraints that separate $H_{\mathrm D}$ from the anisotropy fields. We find that $H_{\mathrm D}$ is weakly temperature dependent, whereas $H_{\mathrm{K1}}$ softens much more rapidly than $H_{\mathrm{K2}}$. This disparity drives the effective uniaxial field to zero at the Morin transition, providing a quantitative microscopic account of the spin reorientation. 
Furthermore, by tracking both $k=0$ and finite-$k$ magnons across the Morin and spin-flop transitions, we compare the finite-$k$ group velocity $v_{\mathrm g}$ with the exchange-controlled limiting group velocity $v_0$. Then we relate their distinct behaviors to the underlying magnetic free-energy parameters and magnon dynamics.
Our results establish low-wavenumber Raman spectroscopy as a powerful probe of finite-momentum antiferromagnetic magnons and provide quantitative guidance for THz magnonic device design.

 \begin{figure}[htb]
	\centering
	\includegraphics[width=\linewidth]{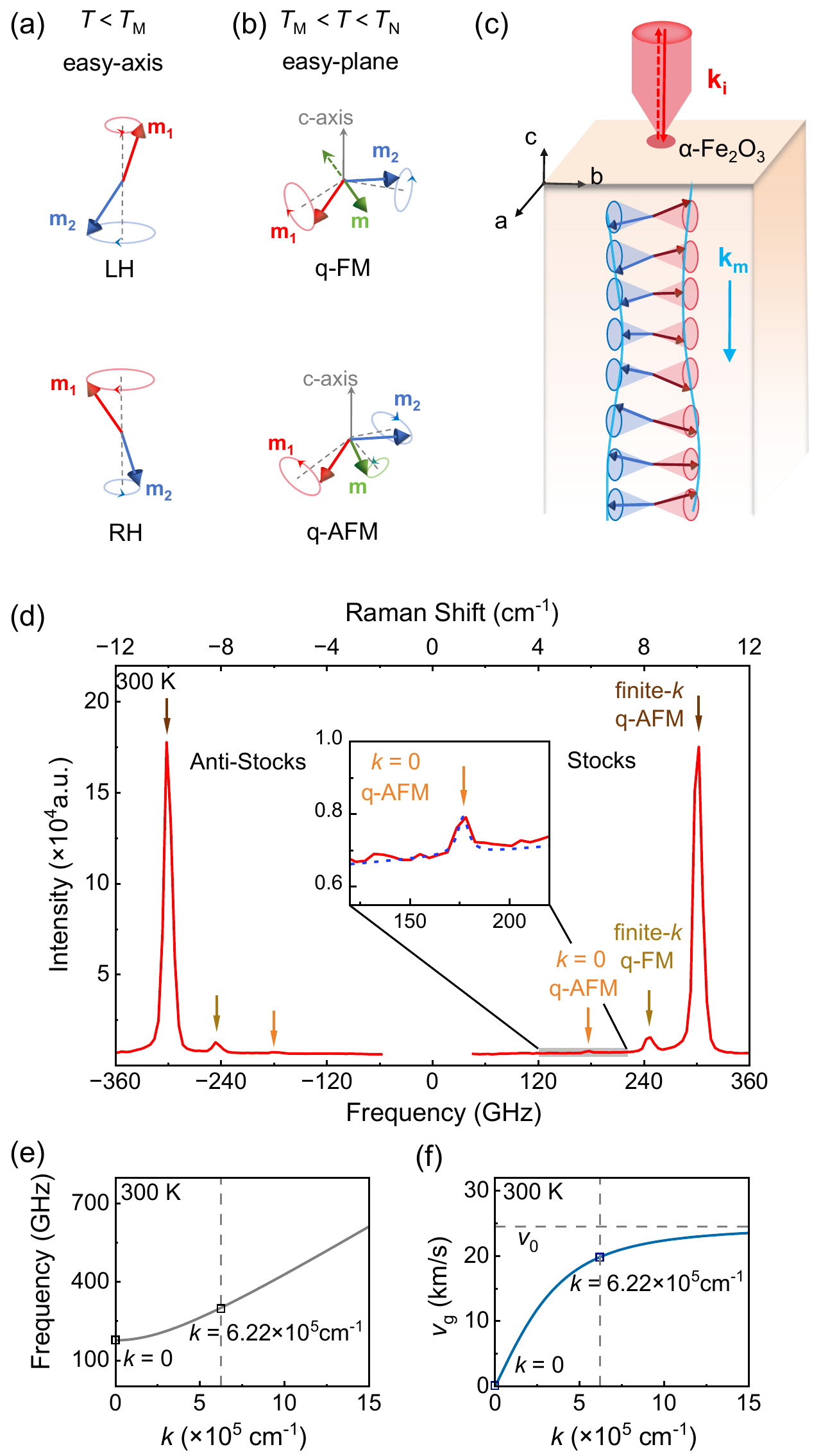}
	\caption{Magnon modes and low-wavenumber Raman measurements of $\alpha$-Fe$_2$O$_3$. (a) Schematic spin precession of Right- and Left-handed magnon modes (RH and LH) in the easy-axis phase at $T<T_{\mathrm{M}}$. (b) Schematic spin precession of q-FM and q-AFM modes in the easy-plane phase at $T>T_{\mathrm{M}}$. (c) Backscattering low-wavenumber Raman geometry for the $c$-cut sample, with $\mathbf{k}_i\parallel\mathbf{c}$ and $\mathbf{k}_m\parallel\mathbf{c}$. (d) Raman spectrum of $\alpha$-Fe$_2$O$_3$ at $300$~K reveals both zone-center and finite-$k$ magnon branches, including $k=0$ q-AFM, finite-$k$ q-FM, and finite-$k$ q-AFM modes. The inset shows an enlarged spectrum of the $k=0$ q-AFM mode.
 (e) Dispersion curve of q-AFM mode at 300~K. (f) Corresponding magnon group velocity $v_{\mathrm{g}}$ as functions of magnon wavevector $k_{\mathrm{m}}$. The measured finite-$k$ modes ($k_{\mathrm{m}}=6.22\times10^5$~cm$^{-1}$) are marked.}
\label{fig:1}
\end{figure}         
                                                                                       
\section{Experiment}
\subsection{Hematite properties and experimental setup}
Hematite is an insulating antiferromagnet with the corundum crystal structure~\cite{pauling1925crystal,zhang2022control,jani2021antiferromagnetic}.
It undergoes the Morin transition at $T_{\mathrm{M}}=255$~K, well below the N\'eel temperature $T_{\mathrm{N}}\sim950$~K~\cite{lafta2020evaluation,majzlan2002thermodynamics}.
For $T < T_{\mathrm{M}}$, the sublattice moments $\mathbf{m}_1$ and $\mathbf{m}_2$ are antiparallel along the crystallographic $c$ axis, forming the easy-axis phase. The two zero-field magnon eigenmodes are the degenerate right-handed (RH) and left-handed (LH) modes, which carry opposite angular momenta~\cite{rezende2019introduction}, as schematically shown in Fig.~\ref{fig:1}(a). For $T > T_{\mathrm{M}}$, the moments reorient into the basal plane. The DM interaction slightly cants the sublattices, producing a weak net moment $\mathbf{m}=\mathbf{m}_1+\mathbf{m}_2$~\cite{wang2021spin}. The two low-energy eigenmodes in this easy-plane phase are the quasi-antiferromagnetic (q-AFM) and quasi-ferromagnetic (q-FM) modes~\cite{han2020birefringence}, as shown in Fig.~\ref{fig:1}(b).

Low-wavenumber Raman scattering measurements were performed in a backscattering geometry on a $0.5$-mm-thick $c$-cut $\alpha\text{-Fe}_2\text{O}_3$ single crystal~\cite{dong2023low}, as shown in Fig.1(c) The sample was mounted in a closed-cycle cryostat equipped with a superconducting magnet ($0\text{--}9$~T). A continuous-wave He-Ne laser (632.8~nm, $500~\mu\text{W}$) was focused along the $c$ axis, and the backscattered light was collected by the same objective. Dictated by momentum conservation, this geometry probes magnons with the wave vector $\mathbf{k}_m \parallel \mathbf{c}$.

A representative zero-field spectrum at 300~K is shown in Fig.~\ref{fig:1}(d). Three distinct peaks emerge and are assigned to the following modes: (1) the $k=0$ q-AFM mode at $177~\text{GHz}$ shown in the insert of Fig.~\ref{fig:1}(d), (2) the finite-$k$ q-FM mode at $246~\text{GHz}$, and (3) the finite-$k$ q-AFM mode at $301~\text{GHz}$. The $k=0$ q-FM mode is not observed as it falls below our detection limit of $\sim 90$~GHz. The corresponding dispersion of q-AFM mode is depicted in Fig.~\ref{fig:1}(e). The wavenumber of the detected magnon, $k_m$, is constrained by the optical Bragg condition~\cite{hortensius2021coherent}:
\begin{equation} k_m = 2k_i n(\lambda)\cos\theta.
\label{eq:1} \end{equation}
where $k_i = 2\pi/\lambda$ is the wavenumber of the incident light in vacuum, $\lambda$ is its vacuum wavelength, $\theta$ is the incident angle and $n$ is the refractive index. Under the condition of normal incidence ($\theta=0$) and utilizing $n = 3.13$ at $\lambda = 632.8$~nm~\cite{querry1985optical}, we obtain $k_{\mathrm{m}} = 6.22 \times 10^5~\text{cm}^{-1}$. 
The finite-$k$ q-AFM frequency follows the long-wavelength antiferromagnetic dispersion~\cite{hortensius2021coherent}:
\begin{equation}
f^2(k_m)
=
f_0^2
+\frac{v_0^2k_m^2}{4\pi^2}.
\label{eq:2}
\end{equation}
where the finite-$k$ magnon frequency $f(k_m)$, the $k=0$ magnon frequency $f_0$, and the limiting magnon group velocity $v_0$ are all temperature-dependent parameters. Using Eqs.~(\ref{eq:1}) and (\ref{eq:2}), we obtained $v_0 = 24.6~\text{km/s}$ and $v_{\mathrm g}(k)=2\pi\frac{\mathrm{d}f(k)}{\mathrm{d}k}$ at 300~K, as shown in Fig.~\ref{fig:1}(f). The resulting $v_{\mathrm{g}} = 19.9~\text{km/s}$ at $k_m = 6.22 \times 10^5~\text{cm}^{-1}$ indicate the potential of hematite for ultra-fast room-temperature magnonic applications.

Furthermore, we measured the excitation-wavelength dependence using 514.4, 532.1, and 632.8~nm lasers. The finite-$k$ magnon intensity is strongly enhanced at 632.8~nm, corresponding to 1.96~eV, close to the $^{6}A_1\rightarrow{}^{4}T_2$ transition of Fe$^{3+}$ ions near 1.98~eV~\cite{sherman1985electronic}. This resonant enhancement, enabled by relaxed selection rules associated with local symmetry breaking and spin-orbit coupling, improves the sensitivity to weak finite-momentum magnon modes.


\begin{figure}[htb]
	\centering
	\includegraphics[width=\linewidth]{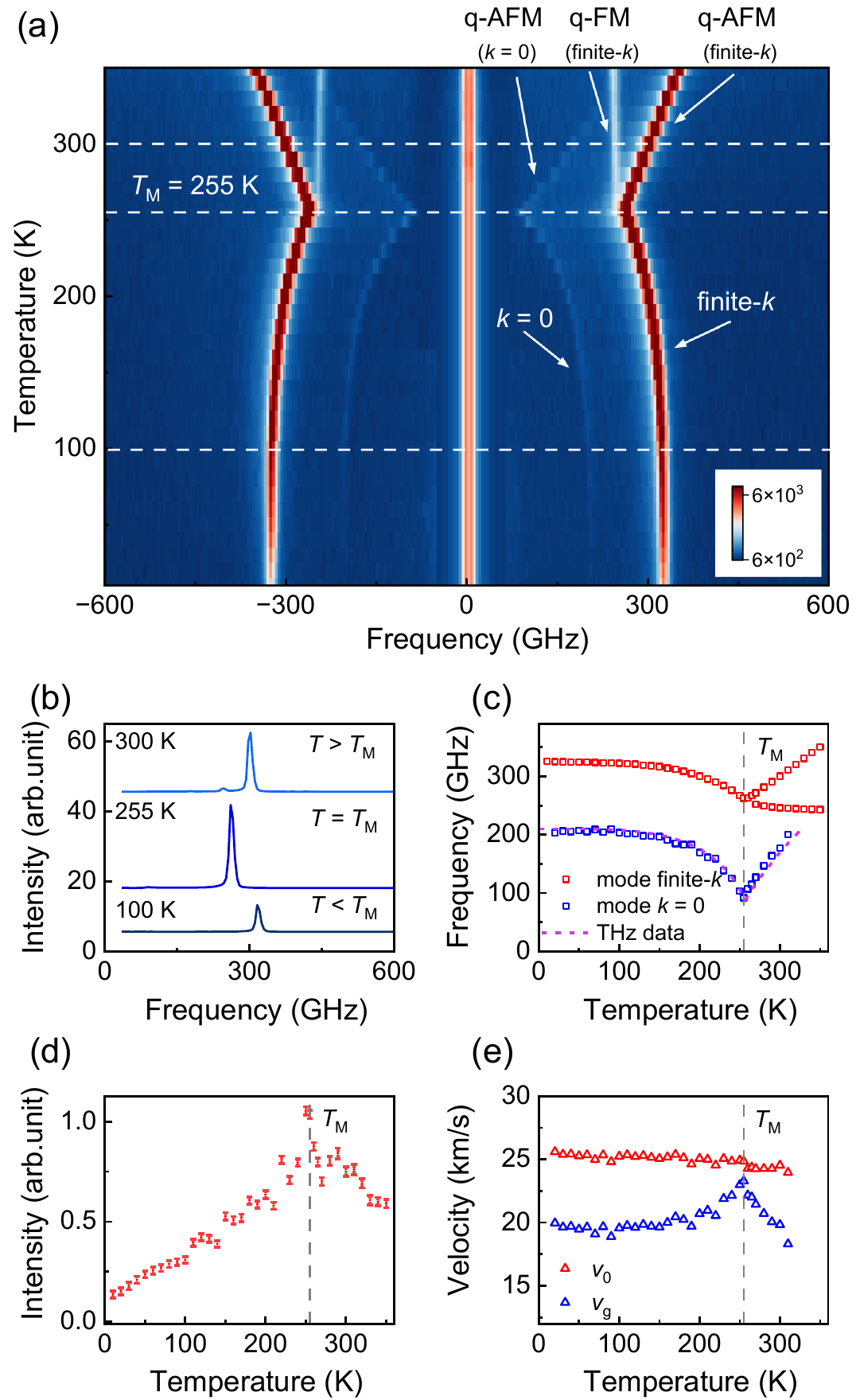}
    \caption{Temperature-dependence of $k=0$ and finite-$k$ magnons at zero field.
    (a) Zero-field Raman spectra measured from 10 to 350~K. The three white-dashed lines mark $100$, $255$, and $300$~K.
    (b) Representative magnon spectra below (100~K), near (255~K), and above (300~K) the Morin temperature $T_{\mathrm{M}}=255$~K.
    (c) Temperature dependence of the extracted $k=0$ (blue squares) and finite-$k$ (red squares) magnon frequencies. The purple dashed line shows the $k=0$ magnon frequency measured by THz spectroscopy~\cite{chou2012high}.
    (d) Temperature dependence of the finite-$k$ magnon Raman intensity. 
    (e) Temperature dependence of the limiting group velocity $v_0$ (red triangles) and the finite-$k$ group velocity $v_{\mathrm{g}}$ (blue triangles). For $T>T_{\mathrm{M}}$, the data in (d) and (e) correspond to the q-AFM mode. Error bars represent one standard deviation. }
    \label{fig:2}
\end{figure}

\subsection{Temperature evolution of finite-momentum magnons}
We next examine the temperature evolution of the Raman-active magnon modes in zero magnetic field. Figure~\ref{fig:2}(a) shows the low-wavenumber Raman spectra measured from 10 to 350~K. Both the zone-center and finite-momentum magnon modes exhibit a pronounced nonmonotonic temperature dependence. To illustrate the evolution more clearly, Fig.~\ref{fig:2}(b) compares representative spectra at 100, 255, and 300~K, corresponding to the easy-axis phase, the vicinity of the Morin transition, and the easy-plane phase, respectively.

At 100~K, the system is in the easy-axis phase. The two circularly polarized antiferromagnetic modes, i.e., the RH and LH branches, are degenerate in zero field and therefore appear as a single peak for both the $k=0$ and finite-$k$ excitations. As the temperature approaches the Morin temperature, both modes soften continuously. At $T_{\mathrm{M}}=255$~K, the $k=0$ and finite-$k$ modes reach their minimum frequencies of approximately 92 and 263~GHz, respectively. This pronounced softening signals the collapse of the anisotropy gap associated with the spin-reorientation transition. As discussed in Sec.~\ref{subsec:afm_parameters}, it originates from the rapid thermal reduction of the first-order uniaxial anisotropy field $H_{\mathrm{K1}}$~\cite{morrish1994canted}.

Above $T_{\mathrm{M}}$, the sublattice moments rotate into the basal plane and the zero-field degeneracy is lifted. The magnon spectrum then separates into the low-frequency q-FM branch and the high-frequency q-AFM branch. At 300~K, the finite-$k$ q-FM and q-AFM modes are clearly resolved at 246 and 301~GHz, respectively. The corresponding zone-center q-AFM mode is observed at 177~GHz, whereas the zone-center q-FM mode lies below the low-frequency cutoff of our setup. The extracted peak positions are summarized in Fig.~\ref{fig:2}(c). The temperature dependence of the $k=0$ mode agrees well with previous THz spectroscopy measurements~\cite{chou2012high}, confirming the reliability of the Raman assignment.

The spectral weight of the finite-$k$ magnon also changes strongly with temperature. Figure~\ref{fig:2}(d) shows the integrated intensity obtained from Lorentzian fits. 
The intensity increases upon approaching $T_{\mathrm{M}}$ from the easy-axis phase and reaches a maximum near the transition. This enhancement can be attributed to the combined effects of increased thermal magnon population and enhanced spin susceptibility near the spin-reorientation transition. The linewidth, characterized by the full width at half maximum, also broadens near $T_{\mathrm{M}}$, indicating enhanced damping and stronger magnetic fluctuations at the phase boundary. Away from the transition, the modes sharpen again as the magnetic order becomes more stable.

Using the measured $k=0$ and finite-$k$ frequencies, we determine the temperature dependence of the limiting group velocity $v_0$ through Eq.~(\ref{eq:2}). As shown in Fig.~\ref{fig:2}(e), $v_0$ decreases gradually from 25.6~km/s at 10~K to 23.9~km/s at 310~K. Because $v_0$ is primarily governed by exchange interactions~\cite{wang2023long}, its slight decrease with temperature suggests a modest thermal weakening of the exchange coupling. In contrast, the group velocity at the Raman-selected finite wave vector, $v_g(k_m)$, displays a pronounced nonmonotonic behavior, consistent with previous reports~\cite{li2026anisotropic,xu2026handed,xu2026optical}. It reaches a maximum value of 23.3~km/s near $T_{\mathrm{M}}$, where the $k=0$ gap is strongly suppressed, and the magnon dispersion becomes steepest at $k_m$. 
Notably, For $T>T_{\mathrm M}$, the integrated intensity and $v_{\mathrm g}$ correspond to the finite-$k$ q-AFM branch, while $v_0$ is extracted from the corresponding q-AFM dispersion using the $k=0$ and finite-$k$ frequencies.

The distinct temperature dependences of $v_0$ and $v_{\mathrm{g}}$ highlight two different aspects of the spin dynamics. The limiting velocity $v_0$ represents the intrinsic exchange-controlled speed scale, whereas the finite-$k$ group velocity is additionally controlled by the anisotropy gap. Consequently, the softening of the magnon gap near the Morin transition enhances $v_{\mathrm{g}}$ even though $v_0$ varies only weakly. The resulting velocity exceeds that of conventional ferromagnetic insulators such as YIG, where typical spin-wave velocities are on the order of 10~km/s~\cite{hamdi2023spin}, and is also substantially larger than values reported for emerging two-dimensional antiferromagnetic semiconductors such as CrSBr, approximately 3--4~km/s~\cite{bae2022exciton}. These results demonstrate that hematite supports exceptionally fast finite-momentum magnons and that their propagation speed can be strongly tuned through the spin-reorientation transition.

\begin{figure}[htb]
	\centering
	\includegraphics[width=\linewidth]{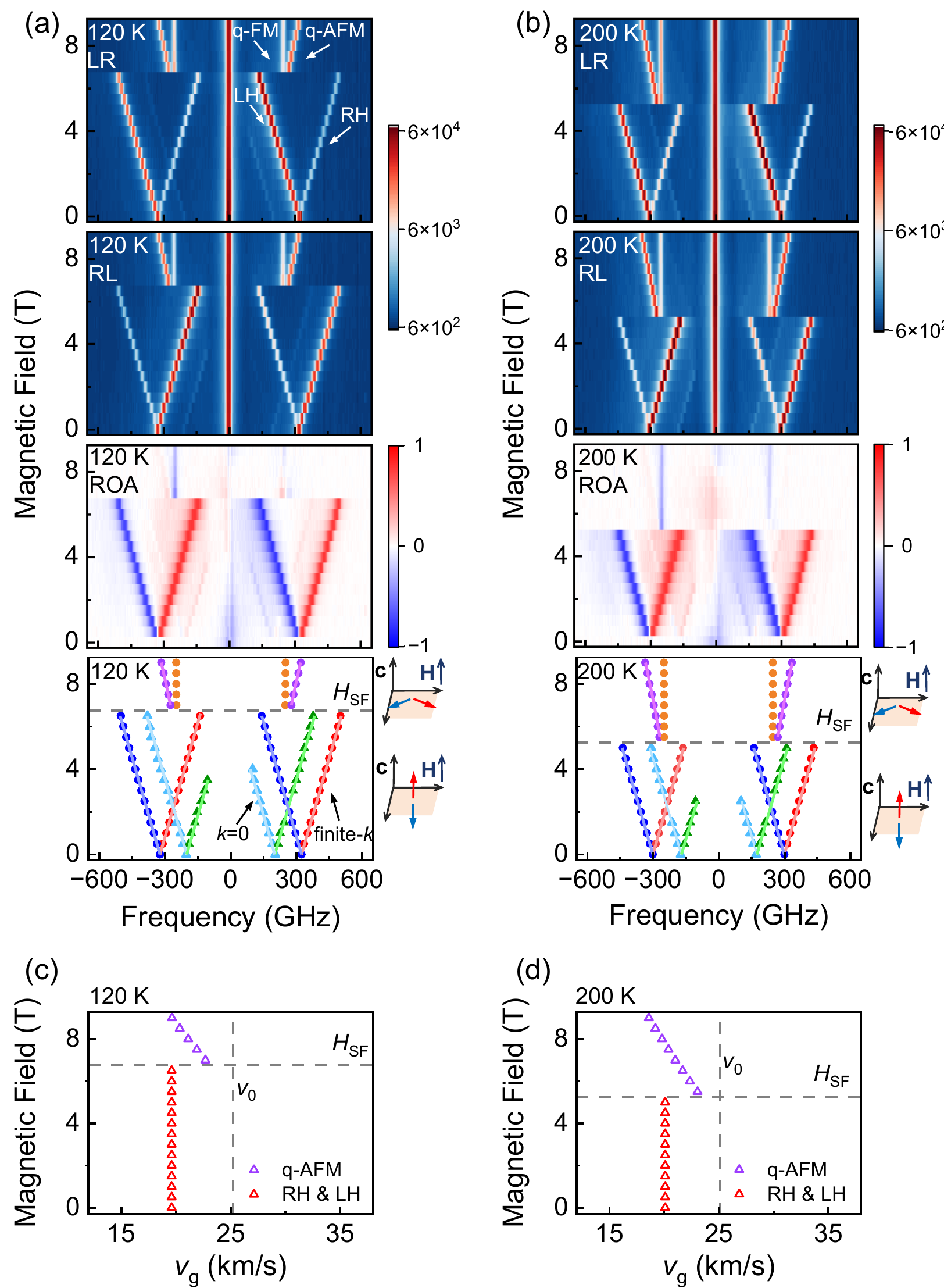}
    \caption{
Field dependence of magnon spectra for $\mathbf{H}\parallel\mathbf{c}$ below $T_{\mathrm{M}}$.
(a),(b) Raman spectra measured in the cross-circular polarization channels RL and LR, together with the corresponding Raman optical activity (ROA), at 120 and 200~K, respectively. The lower panels show the extracted peak positions. Below the spin-flop field $H_{\mathrm{SF}}$, the finite-$k$ right- and left-handed modes are indicated by red and blue circles, respectively, and the $k=0$ right- and left-handed modes by green and blue triangles. The color-coded solid curves are theoretical fits based on Eq.~(\ref{eq:4}). Above $H_{\mathrm{SF}}$, the finite-$k$ q-FM and q-AFM modes are indicated by orange and purple circles, respectively. The solid purple curve is a fit to the q-AFM mode using Eq.~(\ref{eq:5}). The schematics on the right show the spin configurations below and above $H_{\mathrm{SF}}$ for $\mathbf{H}\parallel\mathbf{c}$ at $T<T_{\mathrm M}$.
(c),(d) Magnetic-field dependence of $v_{\mathrm{g}}$ across $H_{\mathrm{SF}}$ at 120 and 200~K, respectively. Dashed lines indicate the field-independent limiting velocities $v_0=25.2$ and $25.1~\mathrm{km/s}$, respectively.
}
    
    \label{fig:3}
\end{figure}

\subsection{Magnetic field dependence of the easy-axis magnons ($T<T_{\mathrm{M}}$)}
\label{subsec:T<TM}
Having established the zero-field temperature evolution of the finite-$k$ magnon spectrum, we now turn to its magnetic-field response in the easy-axis phase, where field-induced mode splitting and spin-flop behavior provide further insight into the underlying magnetic energy landscape. We first investigate the magnetic-field dependence of the easy-axis magnons under an external magnetic field $\mathbf{H} \parallel \mathbf{c}$ up to $9$~T. The magneto-Raman spectra at $120$~K and $200$~K are displayed in Figs.~\ref{fig:3}(a) and \ref{fig:3}(b), respectively, with the parallel field geometry depicted on the right.
To characterize the handedness of the magnon modes, we define the Raman optical activity as $\mathrm{ROA}=(I_{\mathrm{RL}}-I_{\mathrm{LR}})/(I_{\mathrm{RL}}+I_{\mathrm{LR}})$, where $I_{\mathrm{RL}}$ and $I_{\mathrm{LR}}$ are obtained from polarization-resolved measurements in the cross-circular LR and RL channels, corresponding to left-circular excitation/right-circular detection and the reverse configuration, respectively~\cite{wu2023fluctuation,cenker2021direct}.

Below the spin-flop field $H_{\mathrm{SF}}$, the applied field lifts the degeneracy of both the $k=0$ and finite-$k$ magnon modes, producing a lower-frequency LH branch and a higher-frequency RH branch. The opposite signs of their ROA distinguish the angular momenta of the two branches. 
To quantitatively model this Zeeman splitting, we first establish the zero-field $k=0$ magnon frequency~\cite{morrish1994canted}:
\begin{equation}
f_0
=
\frac{\gamma}{2\pi}
\sqrt{
2H_{\mathrm{E}}
\left(
H_{\mathrm{K1}}+2H_{\mathrm{K2}}
\right)
-H_{\mathrm{D}}^2
}.
\label{eq:f0}
\end{equation}
Here, the exchange field is fixed at $H_{\mathrm E}=1040$~T~\cite{lebrun2019anisotropies}. Notably, rather than using a single anisotropy field $H_{\mathrm A}$, we retain $H_{\mathrm{K1}}$ and $H_{\mathrm{K2}}$ separately because $H_{\mathrm{K2}}$ becomes relevant when the Néel vector is reoriented by a transverse field or across the spin-flop transition~\cite{besser1967magnetocrystalline,morrison1973high,dannegger2023magnetic}. Both cases are discussed later.

Under the external field $H$, the split finite-$k$ RH and LH modes follow the linear shift~\cite{morrish1994canted}:
\begin{equation}
f_{\mathrm{RH,LH}}(k_m)
=
\sqrt{
f_0^2
+\frac{v_0^2k_m^2}{4\pi^2}
}
\pm
\frac{\gamma H}{2\pi}.
\label{eq:4}
\end{equation}
Simultaneous fits of the field-dependent frequencies of the $k=0$ and finite-$k$ magnon modes using Eq.~(\ref{eq:4}) yield excellent agreement with the experimental data. Upon increasing the temperature from $120$~K to $200$~K, the Zeeman-split branches display nearly identical slopes, yielding a gyromagnetic ratio of $\gamma/2\pi \simeq 28~\mathrm{GHz/T}$. Furthermore, both $v_{\mathrm{g}}$ and $v_0$ remain unaffected by the external field  as shown in Figs.~\ref{fig:3}(c) and \ref{fig:3}(d), taking values of $v_{\mathrm{g}} = 19.6$ ($20.7$)~km/s and $v_0 = 25.2$ ($25.1$)~km/s at $120~\mathrm{K}$ ($200~\mathrm{K}$). Such velocity stability against magnetic fields holds promise for magnon transport.


Upon reaching $H_{\mathrm{SF}}$, the spins reorient from the $c$-axis into the basal plane. As shown in the high-field regions of Figs.~\ref{fig:3}(a) and \ref{fig:3}(b), $H_{\mathrm{SF}}$ decreases from $6.75$~T at $120$~K to $5.25$~T at $200$~K, consistent with the thermal reduction of the anisotropy field $H_{\mathrm{K1}}$~\cite{morrish1994canted}. Across this spin-flop transition, the magnon spectrum reorganizes into two distinct modes: a nearly field-independent low-frequency q-FM branch and a high-frequency q-AFM branch that blueshifts monotonically with increasing $H$. Notably, their field dependencies closely resemble those of the easy-plane phase in the parallel field configuration and the finite-$k$ q-AFM mode is described by~\cite{morrish1994canted}:
\begin{equation}
f_{\mathrm{qAFM}}^{\parallel}(k_m)
=
\frac{\gamma}{2\pi}
\sqrt{
H^2
+H_{\mathrm{D}}^2
-2H_{\mathrm{E}}H_{\mathrm{K1}}
+\frac{v_0^2k_m^2}{\gamma^2}
}.
\label{eq:5}
\end{equation}
Theoretical fits based on Eq.~(\ref{eq:5}) show excellent agreement with the experimental data, supporting the applicability of the model in the spin-flop phase. Notably, two characteristic changes accompany the spin-flop transition. First, the ROA of the q-AFM mode vanishes, consistent with the theoretically predicted change from circular to purely linear polarization. Second, the calculated magnon velocity $v_{\mathrm{g}}$ exhibits an abrupt change across the field-induced phase boundary, as shown in Figs.~\ref{fig:3}(c) and \ref{fig:3}(d). While $v_0$ remains field-independent, $v_{\mathrm{g}}$ of the q-AFM mode increases discontinuously across $H_{\mathrm{SF}}$, reaching $22.6$ ($23.1$)~km/s at $7$ ($5.5$)~T for $120$~K ($200$~K). As $H$ is further increased to $9$~T, $v_{\mathrm{g}}$ decreases monotonically to $19.6$ ($18.6$)~km/s at $120$~K ($200$~K). This behavior contrasts with the field-independent $v_{\mathrm{g}}$ below $H_{\mathrm{SF}}$ and demonstrates that the spin-flop transition simultaneously reconstructs the polarization and propagation of the finite-$k$ magnons.

\begin{figure}[htb]
	\centering
	\includegraphics[width=\linewidth]{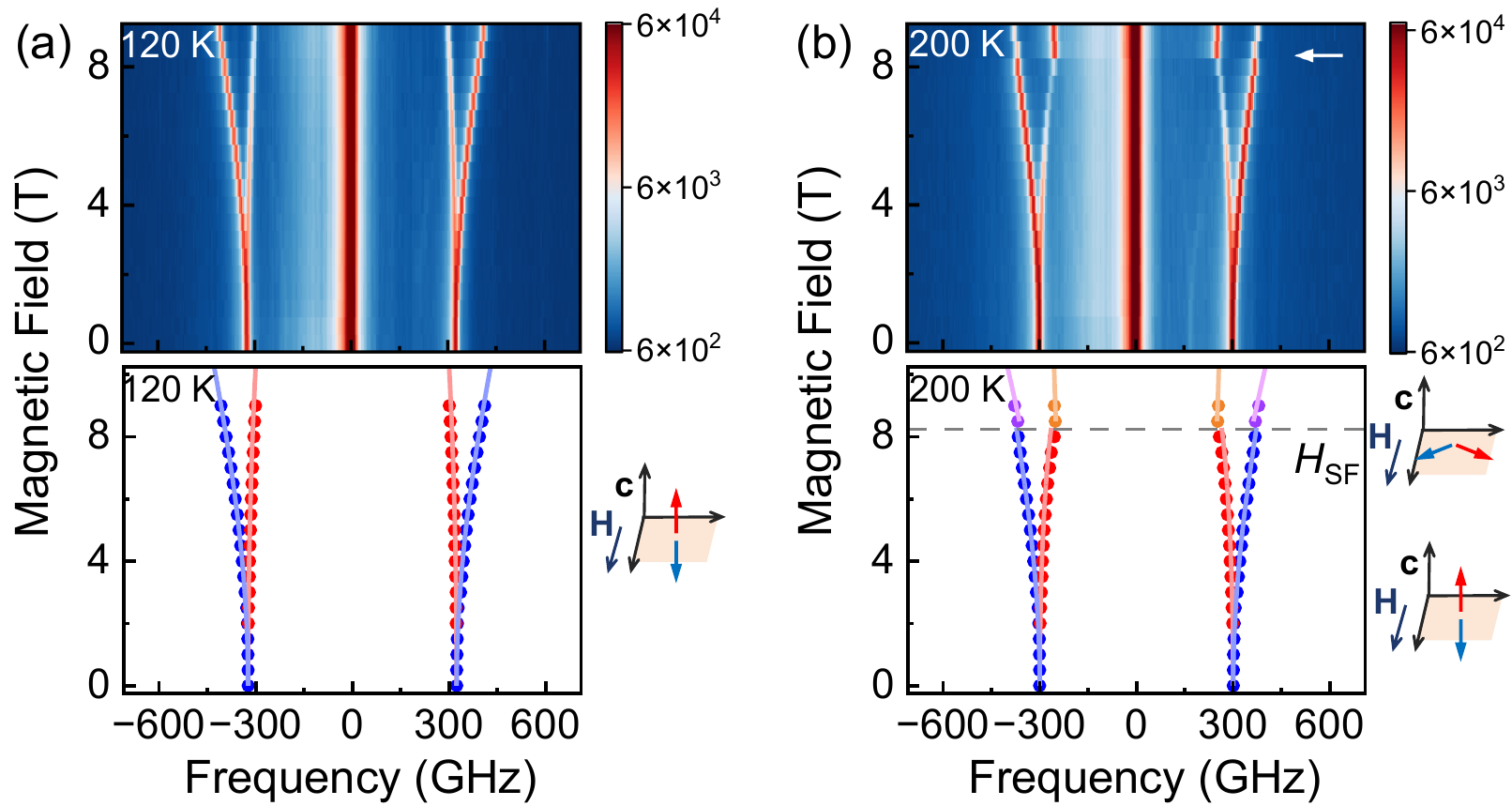}
	\caption{
Field dependence of finite-$k$ magnons for $\mathbf{H}\perp\mathbf{c}$ below $T_{\mathrm{M}}$.
(a),(b) Raman spectra measured at 120 and 200~K, respectively. The lower panels show the extracted finite-$k$ magnon peak positions. Below the spin-flop field $H_{\mathrm{SF}}$, the split finite-$k$ modes are indicated by blue and red circles. In (b), above $H_{\mathrm{SF}}$, the finite-$k$ q-FM and q-AFM modes are indicated by orange and purple circles, respectively. Solid curves are theoretical fits using the expressions derived in Appendix~\ref{app:easy_axis}. The schematics on the right show the spin configurations below and above $H_{\mathrm{SF}}$ for $\mathbf{H}\perp\mathbf{c}$ at $T<T_{\mathrm M}$.
}

    \label{fig:4}
\end{figure}

We next examine the field response for $\mathbf{H} \perp \mathbf{c}$ at $T<T_{\mathrm{M}}$, 
which provides a complementary view of the easy-axis magnon dynamics under a transverse magnetic field. 
Moreover, the transverse-field geometry supplies the independent constraints needed to disentangle the DM and anisotropy fields. 
In the parallel-field configuration, spectra spanning the spin-flop transition constrain $H_{\mathrm{K2}}$ and the combination $H_{\mathrm D}^{2}-2H_{\mathrm E}H_{\mathrm{K1}}$, but cannot determine $H_{\mathrm D}$ and $H_{\mathrm{K1}}$ separately. With these constraints incorporated into the transverse-field model, the field-induced rotation angle of the N\'eel vector becomes a function of the single remaining unknown $H_{\mathrm D}$. The field evolution of the two pre-spin-flop branches therefore allows $H_{\mathrm D}$ to be determined, after which $H_{\mathrm{K1}}$ can be obtained from the parallel-field constraint.

The field-dependent magnon spectra measured at $120$~K and $200$~K are presented in Figs.~\ref{fig:4}(a) and \ref{fig:4}(b), respectively. The two magnon modes that are degenerate at zero field become spectrally resolved at $3$~T. For $3$~T $<H<H_{\mathrm{SF}}$, their frequency gap increases continuously with $H$. 
While $H_{\mathrm{SF}}$ is not observed at 120~K up to the experimental limit of $9$~T, it emerges at $8.25~\mathrm{T}$ when the temperature is increased to 200~K. The decrease in $H_{\mathrm{SF}}$ with increasing temperature is consistent with previous reports~\cite{nagai1975magnon} and can be attributed to the thermal reduction of the anisotropy field $H_{\mathrm{K1}}$~\cite{morrish1994canted}. At this transition, an abrupt Raman spectral weight reversal is observed between the two magnons. Subsequently, the system enters the spin-flop phase, giving rise to the q-FM and q-AFM modes, whose field dependences resemble those observed in the easy-plane phase under the same $\mathbf H\perp\mathbf c$ configuration. The corresponding dispersion model is presented in Appendix~\ref{app:easy_axis}, and the resulting fits reproduce the experimental field dependences well. 

By globally fitting the field-dependent spectra measured under both $\mathbf{H}\parallel\mathbf{c}$ and $\mathbf{H}\perp\mathbf{c}$ using Eqs.~(\ref{eq:f0}), (\ref{eq:4}), and (\ref{eq:5}), together with the dispersion relation provided in Appendix~\ref{app:easy_axis}, we quantitatively determined the key AFM parameters at the two temperatures, as summarized in Table~\ref{tab:afm_parameters}. As the temperature increases from $120$~K to $200$~K, $H_{\mathrm{K1}}$ decreases markedly from approximately $17.0$~mT to $9.6$~mT, whereas $H_{\mathrm{K2}}$ changes only slightly from $5.2$~mT to $5.0$~mT. Meanwhile, $H_{\mathrm D}$ varies only weakly, remaining within $2.0$--$2.1$~T.

\subsection{Magnetic field dependence of the easy-plane magnons ($T>T_{\mathrm{M}}$)}
\label{subsec:T>TM}
\begin{figure}[htb]
	\centering
	\includegraphics[width=\linewidth]{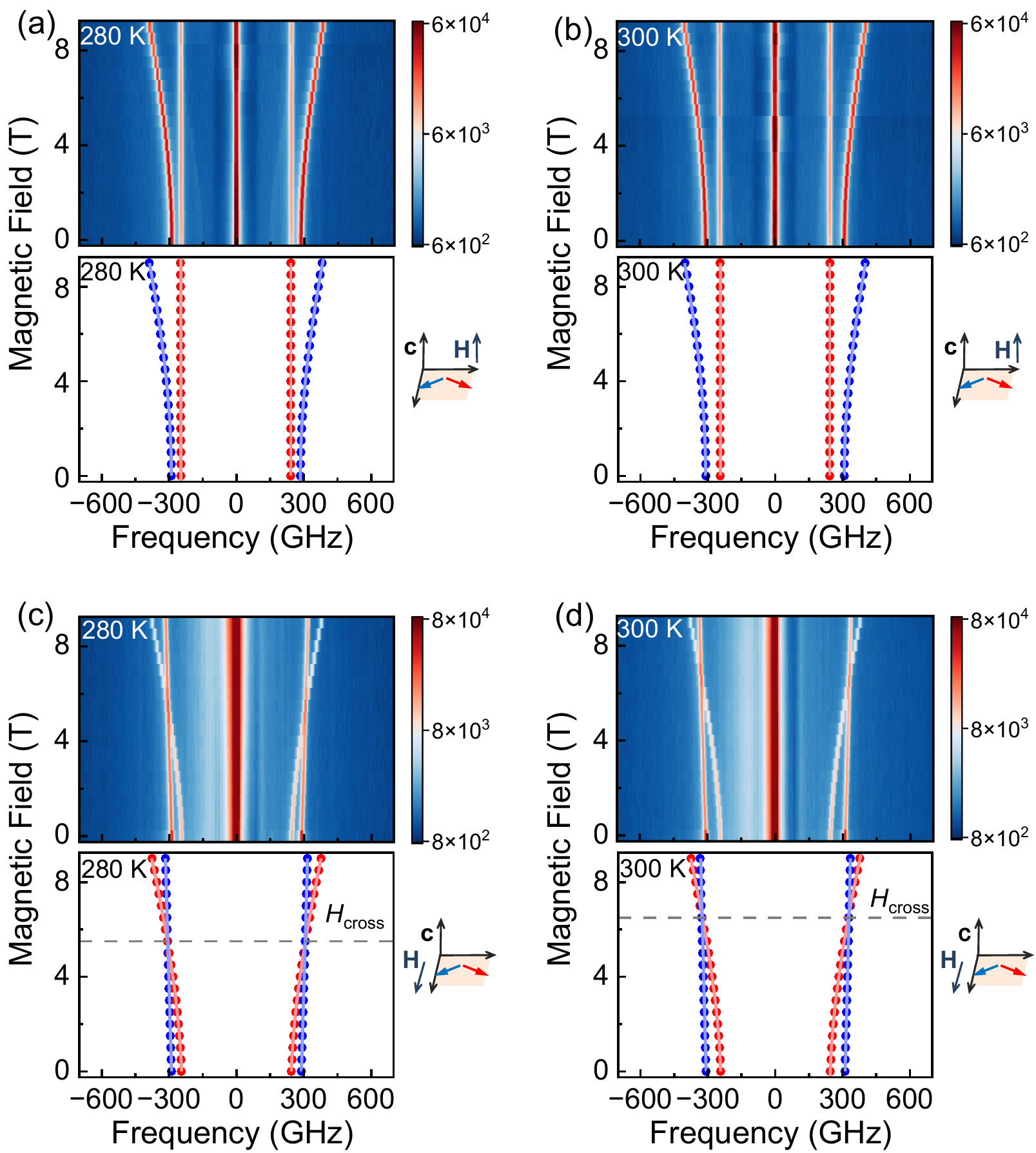}
\caption{
Field dependence of finite-$k$ magnons for $\mathbf{H}\parallel\mathbf{c}$ and $\mathbf{H}\perp\mathbf{c}$ above $T_{\mathrm M}$.
(a),(b) Raman spectra and extracted finite-$k$ magnon peak positions for $\mathbf{H}\parallel\mathbf{c}$ at 280 and 300~K, respectively.
(c),(d) Corresponding spectra and extracted peak positions for $\mathbf{H}\perp\mathbf{c}$ at 280 and 300~K, respectively.
In each panel, the extracted peak positions are shown in the lower subpanel. The q-FM and q-AFM modes are indicated by red and blue circles, respectively. Solid curves are theoretical fits using the expressions derived in Appendix~\ref{app:easy_plane}. The schematics on the right show the magnetic configurations for $\mathbf{H}\parallel\mathbf{c}$ and $\mathbf{H}\perp\mathbf{c}$ at $T>T_{\mathrm M}$.
}
    \label{fig:5}
\end{figure}

We next turn to the easy-plane phase. Figure~\ref{fig:5} summarizes the field dependence of the finite-$k$ magnons measured at $280$~K and $300$~K under both $\mathbf{H}\parallel\mathbf{c}$ and $\mathbf{H}\perp\mathbf{c}$ up to $9$~T. As discussed above, the equilibrium Néel vector lies in the basal plane in both the temperature-induced easy-plane phase and the field-induced spin-flop phase. Consequently, for a given field orientation, these two phases exhibit the same form of finite-$k$ q-AFM and q-FM modes field dispersion. 

At zero field, warming from $280$~K to $300$~K soften the finite-$k$ q-FM mode from $249$ to $246$~GHz, and hardens the q-AFM mode from $282$ to $301$~GHz. Consequently, their zero-field frequency gap widens from $33$ to $55$~GHz. For $\mathbf{H} \parallel \mathbf{c}$, the q-FM mode remains nearly field-independent, while the q-AFM mode blueshifts monotonically. For $\mathbf{H} \perp \mathbf{c}$, in contrast, the q-FM mode rises rapidly in frequency and crosses the q-AFM mode, as presented in Figs.~\ref{fig:5}(c) and (d). The larger zero-field gap at 300~K consequently shifts the crossing field from $5.5$~T at $280$~K to $6.5$~T at $300$~K. 

The easy-plane field dispersion relations and fitting details for easy-plane magnon modes are provided in Appendix~\ref{app:easy_plane}. The fitted curves closely reproduce the peak positions in the lower panels of  of Fig.~\ref{fig:5}. Global fits for both field orientations show that $H_{\mathrm{D}}$ decreases modestly from approximately $2.3$~T at $280$~K to $2.0$~T at $300$~K, whereas $H_{\mathrm{K1}}$ becomes substantially more negative, changing from $-10.2$~mT to $-16.2$~mT, as summarized in Table~\ref{tab:afm_parameters}.

\subsection{Temperature evolution of AFM parameters}
\label{subsec:afm_parameters}
\begin{figure}
    \centering
	\includegraphics[width=\linewidth]{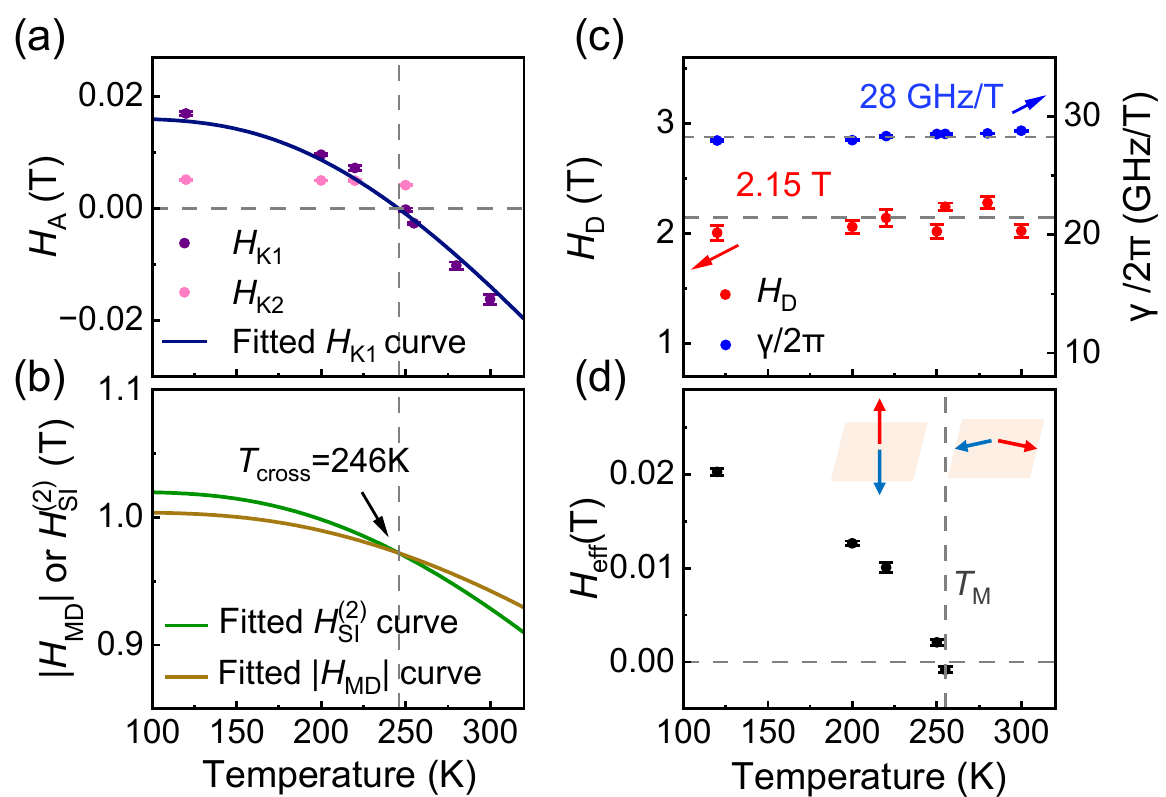}
	\caption{Temperature dependence of the extracted magnetic free-energy parameters. (a) $H_{\mathrm{K1}}$ and $H_{\mathrm{K2}}$. The solid curve represents a fit to $H_{\mathrm{K1}}$. (b) Fitted curves for $H_{\mathrm{SI}}^{(2)}$ and $H_{\mathrm{MD}}$. (c) $H_{\mathrm{D}}$ and $\gamma/2\pi$. (d) $H_{\mathrm{eff}}$ calculated from the extracted parameters. Error bars represent one standard deviation.}
    \label{fig:6}
\end{figure}

In Secs.~\ref{subsec:T<TM} and \ref{subsec:T>TM}, we jointly fitted the field-dependent magnon dispersions measured in the parallel and transverse configurations. This analysis yields $H_{\mathrm D}$, $H_{\mathrm{K1}}$, $H_{\mathrm{K2}}$, and $\gamma/2\pi$ at each temperature. Here we summarize their evolution from 120 to 300~K and discuss the magnetic parameters governing the Morin transition, as shown in Fig.~\ref{fig:6}.

Figure~\ref{fig:6}(a) shows that $H_{\mathrm{K1}}$ decreases sharply from 17.0~mT at 120~K to $-16.2$~mT at 300~K. To describe this behavior,  we use the Besser model~\cite{artman1965magnetic, morrish1994canted,besser1967magnetocrystalline}, in which $H_{\mathrm{K1}}$ is the sum of the second-order single-ion anisotropy $H_{\mathrm{SI}}^{(2)}$ and the magnetic dipole interaction $H_{\mathrm{MD}}$:
\begin{equation}
H_{\mathrm{K1}}(T)= H_{\mathrm{SI}}^{(2)}(T) + H_{\mathrm{MD}}(T)
\label{eq:HK1}
\end{equation}
Here $T$ is the temperature. Within the molecular-field approximation~\cite{besser1967magnetocrystalline}, the temperature dependencies of $H_{\mathrm{MD}}(T)$ and $H_{\mathrm{SI}}^{(2)}(T)$ are provided in Appendix~\ref{app:Molecular-Field Approximation} and plotted in Fig.~\ref{fig:6}(b). Since $H_{\mathrm{MD}}(0)$ is determined via geometrical lattice summation~\cite{besser1967magnetocrystalline}, we fix $H_{\mathrm{MD}}(0) = -1004$~mT~\cite{artman1965magnetic}, and fit the measured $H_{\mathrm{K1}}(T)$ using $H_{\mathrm{SI}}^{(2)}(0)$ as as an tunable parameter. The fit gives $H_{\mathrm{SI}}^{(2)}(0)=1020$~mT and reproduces the measured $H_{\mathrm{K1}}(T)$ well. The result shows that the positive single-ion contribution decreases faster with temperature than $|H_{\mathrm{MD}}|$, driving $H_{\mathrm{K1}}$ through zero at $T_{\mathrm{cross}}=246$~K.

By contrast, $H_{\mathrm{K2}}$ originates from the fourth-order single-ion anisotropy $H_{\mathrm{SI}}^{(4)}$~\cite{morrish1994canted}, which is much smaller than $H_{\mathrm{SI}}^{(2)}$ at low temperature~\cite{dannegger2023magnetic,ovchinnikov2010temperature}. Because the measured temperature range is far below $T_{\mathrm N}\sim950$~K, the normalized magnetization remains nearly saturated, limiting the temperature variation of $H_{\mathrm{SI}}^{(4)}$~\cite{callen1966present}. This accounts for the weak decrease of $H_{\mathrm{K2}}$ from 5.2~mT at 120~K to 4.2~mT at 250~K, as shown in Fig.~\ref{fig:6}(a).

We next consider the Dzyaloshinskii--Moriya field $H_{\mathrm D}$. Previous studies reported values ranging from 2.1 to 4~T using macroscopic magnetic measurements, spin-Hall magnetoresistance, and THz interferometry~\cite{morrish1994canted,lebrun2019anisotropies,liu2023dzyaloshinskii}; however, its continuous temperature evolution across the Morin transition has remained less constrained. Our joint analysis of the parallel- and transverse-field spectra tracks $H_{\mathrm D}$ across both magnetic phases. As shown in Fig.~\ref{fig:6}(c), $H_{\mathrm D}$ remains within 2.0--2.3~T between 120 and 300~K, with a weighted mean of 2.15~T. This weak temperature dependence reflects the robustness of the underlying DMI, which originates from spin-orbit-coupled antisymmetric exchange through the Fe-O-Fe superexchange pathways~\cite{moriya1960anisotropic}. The modest variation in $H_{\mathrm{D}}$ is consistent with limited changes in the local exchange geometry. 

Finally, we reveal that the rapid reduction in $H_{\mathrm{K1}}$ primarily causes the effective anisotropy field $H_{\mathrm{eff}}=H_{\mathrm{K1}} + H_{\mathrm{K2}} - H_{\mathrm{D}}^2/2H_{\mathrm{E}}$~\cite{chow1974soft,ovchinnikov2010temperature,morrish1994canted} to change sign near $T_{\mathrm{M}} = 255$~K, which signal the spin reorientation from the easy-axis to the easy-plane phase, as shown in Fig.~\ref{fig:6}(d). Although $H_{\mathrm{K1}}$ crosses zero at a slightly lower temperature, the transition is governed by the full $H_{\mathrm{eff}}$. Our results identify the rapid softening of $H_{\mathrm{K1}}$ as the main origin of the Morin transition. The extracted $\gamma/2\pi$ remains close to 28~GHz/T over the measured temperature range, corresponding to $g\sim2$, as expected for the nearly spin-only response of high-spin $3d^5$ Fe$^{3+}$ ions ($S=5/2$)~\cite{chou2012high}.

\section{Conclusion}
In conclusion, using low-wavenumber Raman spectroscopy, we tracked the $k=0$ and finite-$k$ magnons in $\alpha\text{-Fe}_2\text{O}_3$ across temperatures from $10$ to $350$~K and magnetic fields up to $9$~T. The simultaneous resolution of these modes allowed us to obtain the exchange-controlled limiting velocity $v_0$ and the finite-$k$ group velocity $v_{\mathrm g}$, revealing their distinct behaviors across the Morin transition and spin-flop transition: $v_{\mathrm g}$ is strongly modified near both transitions, whereas $v_0$ evolves smoothly with temperature and remains nearly field independent. 
Furthermore, by modeling the $\omega$-$H$ dispersion measured in the parallel and transverse field configurations at varying temperatures, we quantitatively extracted and analyzed the temperature evolution of the intrinsic magnetic parameters: $H_{\mathrm{D}}$, $H_{\mathrm{K1}}$, and $H_{\mathrm{K2}}$.
Our results reveal that such a rapid decrease in $H_{\mathrm{K1}}$ drives the sign reversal of the effective anisotropy field $H_{\mathrm{eff}}$ near $T_{\mathrm{M}}=255$~K.
These results provide a quantitative set of magnetic parameters for hematite and demonstrate low-wavenumber magneto-Raman spectroscopy as a sensitive probe of finite-momentum antiferromagnetic magnons in spintronic materials.

\section{Acknowledgements}
The authors thank Prof. Yuan Wan and Prof. Ran Cheng for valuable discussions. This work was supported by the National Key Research and Development Program of China (Grant Nos.~2020YFA0309200), the National Natural Science Foundation of China (Nos.~12474475, 12334007, 12474233), and the Natural Science Foundation of Jiangsu Province (Nos. BK20240057, BK20243011, BK20233001).

\appendix


\section{Field-Dependent Magnon Dispersion for Low-temperature easy-axis phase ($\mathbf{H} \perp \mathbf{c}$)}
\label{app:easy_axis}
In this section, we fit the field-dependent dispersion relations for the magnon branches in the low-temperature easy-axis phase for $\mathbf{H} \perp \mathbf{c}$. Here, we neglect the relatively low in-plane anisotropy according to~\cite{li2026anisotropic}.

The canting angle $\nu$ between the magnetic moments and the $c$ axis is determined by~\cite{morrish1994canted}:
\begin{equation}
\sin\nu = \frac{H H_{\mathrm{D}}}{2H_{\mathrm{E}}(H_{\mathrm{K1}} + 2H_{\mathrm{K2}}\cos^2\nu) - H_{\mathrm{D}}^2}
\label{eq:A1}
\end{equation}

For $H < H_{\mathrm{SF}}$, The corresponding magnon frequencies for the finite-$k$ split branches, denoted as the upper [$f_{+}^{\perp}(k_m)$] and lower [$f_{-}^{\perp}(k_m)$] modes, are~\cite{morrish1994canted}:

\begin{equation}
f_{+}^{\perp}(k_m)
=
\sqrt{
f_0^2
+
\frac{
\gamma^2
\left(
H^2-4H_{\mathrm{E}}H_{\mathrm{K2}}\sin^2\nu
\right)
+v_0^2k_m^2
}{4\pi^2}
}.
\label{eq:A2}
\end{equation}

\begin{equation}
f_{-}^{\perp}(k_m)
=
\sqrt{
f_0^2\cos^2\nu
+
\frac{
v_0^2k_m^2
-3\gamma^2H_{\mathrm{E}}H_{\mathrm{K2}}
\sin^2(2\nu)
}{4\pi^2}
}.
\label{eq:A3}
\end{equation}


For $H > H_{\mathrm{SF}}$, the magnon frequencies for the finite-$k$ q-FM [$f_{\mathrm{qFM}}^{\perp}(k_m)$] and q-AFM [$f_{\mathrm{qAFM}}^{\perp}(k_m)$] modes are expressed as~\cite{morrish1994canted}:

\begin{equation}
f_{\mathrm{qFM}}^{\perp}(k_m)
=
\frac{\gamma}{2\pi}
\sqrt{
H_{\mathrm{D}}\left(H_{\mathrm{D}}+H\right)
-2H_{\mathrm{E}}H_{\mathrm{K1}}
+\frac{v_0^2k_m^2}{\gamma^2}
}.
\label{eq:A4}
\end{equation}

\begin{equation}
f_{\mathrm{qAFM}}^{\perp}(k_m)
=
\frac{\gamma}{2\pi}
\sqrt{
H\left(H+H_{\mathrm{D}}\right)
+\frac{v_0^2k_m^2}{\gamma^2}
}.
\label{eq:A5}
\end{equation}


\section{Field-Dependent Magnon Dispersion for High-temperature easy-plane phase ($\mathbf{H} \parallel \mathbf{c}$ and $\mathbf{H} \perp \mathbf{c}$)}
\label{app:easy_plane}
In this section, we fit the field-dependent dispersion relations for the magnon branches in both the high-temperature easy-plane phase for $\mathbf{H} \parallel \mathbf{c}$ and $\mathbf{H} \perp \mathbf{c}$. Here, we neglect the relatively low in-plane anisotropy according to~\cite{li2026anisotropic}.
First, we define the zero-field baseline frequencies for the q-FM [$f_{\mathrm{qFM}}(k_m)$] and q-AFM [$f_{\mathrm{qAFM}}(k_m)$] magnons with $k_m$ as~\cite{morrish1994canted}:
\begin{equation}
f_{\mathrm{qFM}}(k_m)
=
\frac{v_0k_m}{2\pi}.
\label{eq:qfm0}
\end{equation}

\begin{equation}
f_{\mathrm{qAFM}}(k_m)
=
\frac{\gamma}{2\pi}
\sqrt{
H_{\mathrm{D}}^2
-2H_{\mathrm{E}}H_{\mathrm{K1}}
+\frac{v_0^2k_m^2}{\gamma^2}
}.
\label{eq:qafm0}
\end{equation}

\subsection{Configuration $\mathbf{H} \parallel \mathbf{c}$:}
The field-dependent magnon frequencies for the q-FM [$f_{\mathrm{qFM}}^{\parallel}(k_m)$] and q-AFM [$f_{\mathrm{qAFM}}^{\parallel}(k_m)$] modes are~\cite{morrish1994canted}:

\begin{equation}
f_{\mathrm{qFM}}^{\parallel}(k_m)
=
f_{\mathrm{qFM}}(k_m).
\label{eq:A6}
\end{equation}

\begin{equation}
f_{\mathrm{qAFM}}^{\parallel}(k_m)
=
\sqrt{
f_{\mathrm{qAFM}}^2(k_m)
+\frac{\gamma^2H^2}{4\pi^2}
}.
\label{eq:A7}
\end{equation}


\subsection{Configuration $\mathbf{H} \perp \mathbf{c}$:}



The field-dependent magnon frequencies for the q-FM
($f_{\mathrm{qFM}}^{\perp}(k_m)$) and q-AFM
($f_{\mathrm{qAFM}}^{\perp}(k_m)$) modes are~\cite{morrish1994canted}:
\par
\vspace*{10pt}

\begin{align}
f_{\mathrm{qFM}}^{\perp}(k_m)
&=
\sqrt{
f_{\mathrm{qFM}}^2(k_m)
+
\frac{\gamma^2H(H+H_{\mathrm D})}{4\pi^2}
},
\label{eq:qFM_perp}
\\[10pt]
f_{\mathrm{qAFM}}^{\perp}(k_m)
&=
\sqrt{
f_{\mathrm{qAFM}}^2(k_m)
+
\frac{\gamma^2HH_{\mathrm D}}{4\pi^2}
}.
\label{eq:qAFM_perp}
\end{align}

\section{Molecular-Field Approximation}
\label{app:Molecular-Field Approximation}
In this section, we fitted $H_{\mathrm{MD}}(T)$ and $H_{\mathrm{SI}}^{(2)}(T)$, which are modeled within the molecular-field approximation~\cite{besser1967magnetocrystalline}:
\begin{equation}
H_{\mathrm{MD}}(T)
=
H_{\mathrm{MD}}(0)B_S^2(x).
\label{eq:HMD_temperature}
\end{equation}

\begin{equation}
H_{\mathrm{SI}}^{(2)}(T)
=
\frac{H_{\mathrm{SI}}^{(2)}(0)}{2S-1}
\Biggl[
2(S+1)
-3B_S(x)\coth\left(\frac{x}{2S}\right)
\Biggr].
\label{eq:HSI2_temperature}
\end{equation}

where \(B_S(x) = \frac{2S + 1}{2S} \coth \left( \frac{2S + 1}{2S} x \right) - \frac{1}{2S} \coth \left( \frac{x}{2S} \right)\) is the Brillouin function for $S=5/2$, and the variable $x$ satisfies the self-consistent relation $x = \frac{3S}{S+1} \frac{T_N}{T} B_S(x)$. $H_{\mathrm{MD}}(0)$ and $H_{\mathrm{SI}}^{(2)}(0)$ are $H_{\mathrm{MD}}$ and $H_{\mathrm{SI}}^{(2)}$ at $T=0$~K, respectively.

\section{AFM Parameters}
Table~\ref{tab:afm_parameters} summarizes representative magnetic parameters and magnon velocities at selected temperatures. The data at $120$ and $200$~K correspond to the easy-axis phase, those at $250$ and 255~K to the vicinity of the Morin transition, and those at $280$ and $300$~K to the easy-plane phase. The fields $H_{\mathrm{K1}}$, $H_{\mathrm{K2}}$, and $H_{\mathrm D}$ were extracted from joint fits to the field-dependent magnon spectra, whereas $v_0$ and $v_{\mathrm g}$ were determined from the $k=0$ and finite-$k$ magnon frequencies.
\begin{table}[H]
    \caption{Extracted parameters at selected temperatures for $k_m=6.22\times10^5$~cm$^{-1}$.}
    \label{tab:afm_parameters}
    \centering
    \renewcommand{\arraystretch}{1.15}

    \begin{tabular*}{\columnwidth}
        {@{\extracolsep{\fill}}cccccc@{}}
        \toprule
        $T$ &
        $H_{\mathrm{K1}}$ &
        $H_{\mathrm{K2}}$ &
        $H_{\mathrm D}$ &
        $v_{\mathrm g}$ &
        $v_0$ \\
        (K) & (mT) & (mT) & (T) & (km/s) & (km/s) \\
        \midrule
        $120$ & $17.0$  & $5.2$ & $2.0$ & $19.6$ & $25.2$ \\
        $200$ & $9.6$  & $5.0$ & $2.1$ & $20.7$ & $25.1$ \\
        $250$ & $-0.2$ & $4.2$ & $2.0$ & $23.0$ & $24.9$ \\
        $255$ & $-2.6$  & --     & $2.2$ & $23.3$ & $24.9$ \\
        $280$ & $-10.2$ & --    & $2.3$ & $20.7$ & $24.3$ \\
        $300$ & $-16.2$ & --    & $2.0$ & $19.9$ & $24.6$ \\

        \bottomrule
    \end{tabular*}
\end{table}

\end{document}